\documentclass[preprints,article,accept,oneauthor,pdftex,10pt,a4paper]{mdpi} 

\firstpage{1} 
\pubvolume{xx}
\issuenum{1}
\articlenumber{1}
\pubyear{2018}
\copyrightyear{2018}
\history{}

\pdfoutput=1

\Title{A New Class of Retrocausal Models}

\Author{Ken Wharton}

\AuthorNames{Ken Wharton}

\address[1]{%
\quad Department of Physics and Astronomy, San Jos\'e State University, San Jos\'e, CA 95192-0106 }

\corres{Correspondence: kenneth.wharton@sjsu.edu}

\abstract{Globally-constrained classical fields provide a unexplored framework for modeling quantum phenomena, including apparent particle-like behavior.  By allowing controllable constraints on unknown past fields, these models are retrocausal but not retro-signaling, respecting the conventional block universe viewpoint of classical spacetime.  Several example models are developed that resolve the most essential problems with using classical electromagnetic fields to explain single-photon phenomena.  These models share some similarities with Stochastic Electrodynamics, but without the infinite background energy problem, and with a clear path to explaining entanglement phenomena.  Intriguingly, the average intermediate field intensities share a surprising connection with quantum ``weak values'', even in the single-photon limit.  This new class of models is hoped to guide further research into spacetime-based accounts of weak values, entanglement, and other quantum phenomena.}

\newcommand{\ket}[1]{{|{#1}\!\!>}}
\newcommand{\bra}[1]{{<\!\!{#1}|}}
\newcommand{\ip}[2]{{<\!\!{#1}|{#2}\!\!>}}

\begin{document}



\section{Introduction}

In principle, retrocausal models of quantum phenomena offer the enticing possibility of replacing the high-dimensional configuration space of quantum mechanics with ordinary spacetime, without breaking Lorentz covariance or utilizing action-at-a-distance.\cite{sutherland1983,price1997,wharton2014,price2015a,leifer2017a,adlam2018}.   Any quantum model based entirely on spacetime-localized parameters would obviously be much easier to reconcile with general relativity, not to mention macroscopic classical observations.\footnote{Block-universe retrocausal models can generally violate Bell-type inequalities because they contain hidden variables $\lambda$ that are constrained by the future measurement settings ($a,b$).  These constraints can be mediated via continuous influence on the particle worldlines, explicitly violating the independence assumption $P(\lambda|a,b)=P(\lambda)$ utilized in Bell-type no-go theorems.} 

In practice, however, the most sophisticated spacetime-based retrocausal models to date only apply to a pair of maximally entangled particles.\cite{argaman2010,wharton2014,almada2015,weinstein2017}  A recent retrocausal proposal from Sen \cite{sen2018} is more likely to extend to more of quantum theory, but without a retrocausal mechanism it would have to use calculations in configuration space, preparing whatever initial distribution is needed to match the expected final measurement.  Sutherland's retrocausal Bohmian model \cite{sutherland2017} also uses some calculations in configuration space (and does not assign outcome probabilities).  Given the difficulties in extending known retrocausal models to more sophisticated situations, further development may require entirely new approaches.

One obvious way to change the character of existing retrocausal models is to replace the usual particle ontology with a framework build upon spacetime-based fields.  Every quantum ``particle'', after all, is thought to actually be an excitation of a quantum field, and every quantum field has a corresponding classical field that could exist in ordinary spacetime.  The classical Dirac field, for example, is a Dirac-spinor-valued function of ordinary spacetime, and is arguably a far closer analog to the electrons of quantum theory than a classical charged particle.  This point is even more obvious when it comes to photons, which have no classical particle analog at all, but of course have a classical analog in the ordinary electromagnetic field.

This paper will outline a new class of field-based retrocausal models.  Field-based accounts of particle phenomena are rare but not unprecedented, one example being the Bohmian account of photons \cite{bohm1987,kaloyerou2006}, using fields in configuration space.  One disadvantage to field-based models is that they are more complicated than particle models.  However, if the reason that particle-based models can't be extended to more realistic situations is that particles are too simple, then moving to the closer analog of classical fields might arguably be beneficial.  Indeed, many quantum phenomena (superposition, interference, importance of relative phases, etc.) have excellent analogs in classical field behavior.  In contrast, particles have essentially only one phenomenological advantage over fields: localized position measurements.  The class of models proposed here may contain a solution to this problem, but the primary goal will be to set up a framework in which more detailed models can be developed (and to show that this framework is consistent with some known experimental results).  

Apart from being an inherently closer analog to standard quantum theory, retrocausal field models have a few other interesting advantages to their particle counterparts.  One intriguing development, outlined in detail below, is an account of the average ``weak values'' \cite{AAV,dressel2014a} measured in actual experiments, naturally emerging from the analysis of the intermediate field values.  Another point of interest is that the framework here bears similarities to Stochastic Electrodynamics (SED), but without some of the conceptual difficulties encountered by that program (\textit{i.e.} infinite background energy, and a lack of a response to Bell's theorem).\cite{SED1,SED2}  Therefore it seems hopeful that many of the successes of SED might be applied to a further development of this framework.

The plan of this paper is to start with a conceptual framework, motivating and explaining the general approach that will be utilized by the specific models.  Section 3 then explores a simple example model that illustrates the general approach, as well as demonstrating how discrete outcomes can still be consistent with a field-based model.  Section 4 then steps back to examine a large class of models, calculating the many-run average predictions given a minimal set of assumptions.  These averages are then shown to essentially match the weak-value measurements.  The results are then used to motivate an improved model, as discussed in Section 5, followed by preliminary conclusions and future research directions.
 
\section{Conceptual Framework}

Classical fields generally have Cauchy data on every spacelike hypersurface.  Specifically, for second order field equations, knowledge of the field and its time derivative everywhere at one time is sufficient to calculate the field at all times.   But the uncertainty principle, applied in a field framework, implies that knowledge of this Cauchy data can never be obtained: No matter how precise a measurement, some components of the field can always elude detection.  Therefore it is impossible to assert that either the preparation or the measurement of a field represents the precise field configuration at that time.  This point sheds serious doubt on the way that preparations are normally treated as exact initial boundary conditions (and, in most retrocausal models, the way that measurements are treated as exact final boundary conditions).

In accordance with this uncertainty, the field of Stochastic Electrodynamics (SED) explores the possibility that in addition to measured electromagnetic (EM) field values, there exists an unknown and unmeasured `classical zero-point' EM field that interacts with charges in the usual manner.\cite{SED1,SED2}  Starting from the assumption of relativistic covariance, a natural gaussian noise spectrum is derived, fixing one free parameter to match the effective quantum zero-point spectrum of a half-photon per EM field mode.  Using classical physics, a remarkable range of quantum phenomena can be recovered from this assumption.  But these SED successes come with two enormous problems.  First, the background spectrum diverges, implying an infinite stress energy tensor at every point in spacetime.  Such a field would clearly be in conflict with our best understanding of general relativity, even with some additional ultraviolet cutoff.  Second, there is no path to recovering \textit{all} quantum phenomena via locally interacting fields, because of Bell-inequality violations in entanglement experiments.   

Both of these problems have a potential resolution when using the Lagrangian Schema \cite{wharton2014} familiar from least-action principles in classical physics.  Instead of treating a spacetime system as a computer program that takes the past as an input and generates the future as an output, the Lagrangian Schema utilizes both past and future constraints, solving for entire spacetime structures ``all at once''.  Unknown past parameters (say, the initial angle of a ray of light constrained by Fermat's principle of least time) are the \textit{outputs} of such a calculation, not inputs.  Crucially, the action $S$ that is utilized by these calculations is a covariant scalar, and therefore provides path to a Lorentz covariant calculation of unknown field parameters, different from the divergent spectrum considered by SED.  The key idea is to keep the action extremized as usual ($\delta S=0$), while also imposing some additional constraint on the total action of the system.\footnote{One intriguing option is to quantize the action ($S=nh$), a successful strategy from the ``old'' quantum theory that has not been pursued in a field context (and would also motivate $\delta S=0$ in the first place).}  (Here the action $S$ is the usual functional of the fields throughout any given spacetime subsystem, calculated by integrating the classical Lagrangian density over spacetime.)  

Constraining the action does not merely ensure relativistic covariance.  When complex macroscopic systems are included in the spacetime subsystem (i.e. preparation and measurement devices), they will obviously dominate the action, acting as enormous constraints on the microscopic fields, just as a thermal reservoir acts as a constraint on a single atom.  The behavior of microscopic fields would therefore depend on what experimental apparatus is considered.  Crucially, the action is an integral over spacetime systems, not merely spatial systems.  Therefore the future settings and orientations of measurement devices strongly influence the total action, and unknown microscopic fields at earlier times will be effectively constrained by those future devices.  Again, those earlier field values are literally ``outputs'' of the full calculation, while the measurement settings are inputs. 

Such models are correctly termed ''retrocausal''.  Given the usual block universe framework from classical field theory and the interventionist definition of causation \cite{woodward2005,price1991,pearl2009,menzies1993}, any devices with free external settings are ``causes'', and any constrained parameters are ``effects'' (including field values at spacetime locations before the settings are chosen).  Such models are retrocausal but not retro-signaling, because the future settings constrain \textit{unknown} past field parameters, hidden by the uncertainty principle.  (These models are also forward-causal, because the preparation is another intervention.)  It is important not to view causation as a process -- certainly not one ``flowing'' back-and-forth through time, as this would violate the block universe perspective.  Instead, such systems are consistently solved ``all-at-once'', as in action principles.  Additional discussion of this topic can be found in  \cite{price1997,price2008,price2015a}.

The retrocausal character of these models immediately provides a potential resolution to both of the problems with SED.  Concerning the infinite-density zero point spectrum, SED assumes that all possible field modes are required because one never knows which ones will be relevant in the future.  But a retrocausal model is not ``in the dark'' about the future, because (in this case) the action is an integral that \textit{includes} the future.  And the total action might very well only be highly sensitive to a bare few field modes.  (Indeed, this is usually the case; consider an excited atom, waiting for a zero-point field to trigger ``spontaneous'' emission.  Here only one particular EM mode is required to explain the eventual emission of a photon, with the rest of the zero point field modes being irrelevant to a future photon detector.)  As will be shown below, it is not difficult to envision action constraints where typically only a few field modes need to be populated in the first place, resolving the problem of infinities encountered by SED.  Furthermore, it is well-known that retrocausal models can naturally resolve Bell-inequality violations without action-at-a-distance, because the past hidden variables are naturally correlated with the future measurement settings. \cite{leifer2014,price2015a} (Numerous proof-of-principle retrocausal models of entanglement phenomena have been developed over the past decade \cite{argaman2010,wharton2014,almada2015,weinstein2017,sen2018}.)

Unfortunately, solving for the exact action of even the simplest experiments is very hard.  The macroscopic nature of preparation and measurement that makes them so potent as boundary constraints also makes them notoriously difficult to calculate exactly -- especially when the relevant changes in the action are on the order of Planck's constant.  Therefore, in order to initially consider such models, this paper will assume that any constraint on the total action manifests itself as certain rules constraining how microscopic fields are allowed to interact with the macroscopic devices.  (Presumably, such rules would include quantization conditions, for example only allowing absorption of EM waves in packets of energy $\hbar\omega$.)  This assumption will allow us to focus on what is happening between devices rather than in the devices themselves, setting aside those difficulties as a topic for future research.

This paper will proceed by simply exploring some possible higher-level interaction constraints (guided by other general principles such as time-symmetry), and determining whether they might plausibly lead to an accurate explanation of observed phenomena.  At this level, the relativistic covariance will not be obvious; after all, when considering intermediate EM fields in a laboratory experiment, a special reference frame is determined by the macroscopic devices which constrain those fields.  But it seems plausible that if some higher-level model matches known experiments then a lower-level covariant account would eventually be acheivable, given that known experiments respect relativistic covariance. 

The following examples will be focused on simple problems, with much attention given to the case where a single photon passes through a beamsplitter and is then measured on one path or the other.  This is precisely the case where field approaches are thought to fail entirely, and therefore the most in need of careful analysis.  Also, bear in mind that these are representative \textit{examples} of an entire class of models, not one particular model.  It is hoped that by laying out this new class of retrocausal models, that one particular model will eventually emerge as a possible basis for a future reformulation of quantum theory.

\section{Constrained Classical Fields}

\subsection{Classical Photons}

Ordinary electromagnetism provides a natural analog to a single photon: a finite-duration electromagnetic wave with total energy $\hbar\omega$.  Even in classical physics, all of the usual uncertainty relations exist between the wave's duration and its frequency $\omega$; in the below analysis we will assume long-duration EM waves that have a reasonably well-defined frequency, in some well-defined beam such as the $TEM_{00}$ gaussian mode of a narrow bandwidth laser.  By normalizing the peak intensity $I$ of this wave so that a total energy of $\hbar\omega$ corresponds to $I\!=\!1$, one can define a ``Classical Photon Analog'', or CPA.

Such CPAs are rarely considered, for the simple reason that they seem incompatible with the simple experiment shown in Figure 1a.  If such a CPA were incident upon a beamsplitter, some fraction $T$ of the energy would be transmitted and the remaining fraction $R=1-T$ would be reflected.  This means that detectors $A$ and $B$ on these two paths would never see what actually happens, which is a full $\hbar\omega$ amount of energy on \textit{either} $A$ or $B$, with probabilities $T$ and $R$ respectively.  Indeed, this very experiment is usually viewed as proof that classical EM is incorrect.

Notice that the analysis in the previous paragraph assumed that the initial conditions were exactly known, which would violate the uncertainty principle.  If unknown fields existed on top of the original CPA, boosting its total energy to something larger than $\hbar\omega$, this would change the analysis.  For example, if the CPA resulted from a typical laser, the ultimate source of the photon could be traced back to a spontaneous emission event, and (in SED-style theories) such ``spontaneous'' emission is actually \textit{stimulated} emission, due to unknown incident zero-point radiation.  This unknown background would then still be present, boosting the intensity of the CPA such that $I\!>\!1$.  Furthermore, every beamsplitter has a ``dark'' input port, from which any input radiation would also end up on the same two detectors, $A$ and $B$.  In quantum electrodynamics, it is essential that one remember to put an input vacuum state on such dark ports; the classical analog of this well-known procedure is to allow for possible unknown EM wave inputs from this direction.

The uncertain field strengths apply to the outputs as well as the inputs, from both time-symmetry and the uncertainty principle.  Just because a CPA is measured on some detector $A$, it does not follow that there is no additional EM wave energy that goes unmeasured.  And just because nothing is measured on detector $B$ does not mean that there is no EM wave energy there at all.  If one were to insist on a perfectly energy-free detector, one would violate the uncertainty principle.

\begin{figure}[H]
\centering
\includegraphics[width=10 cm]{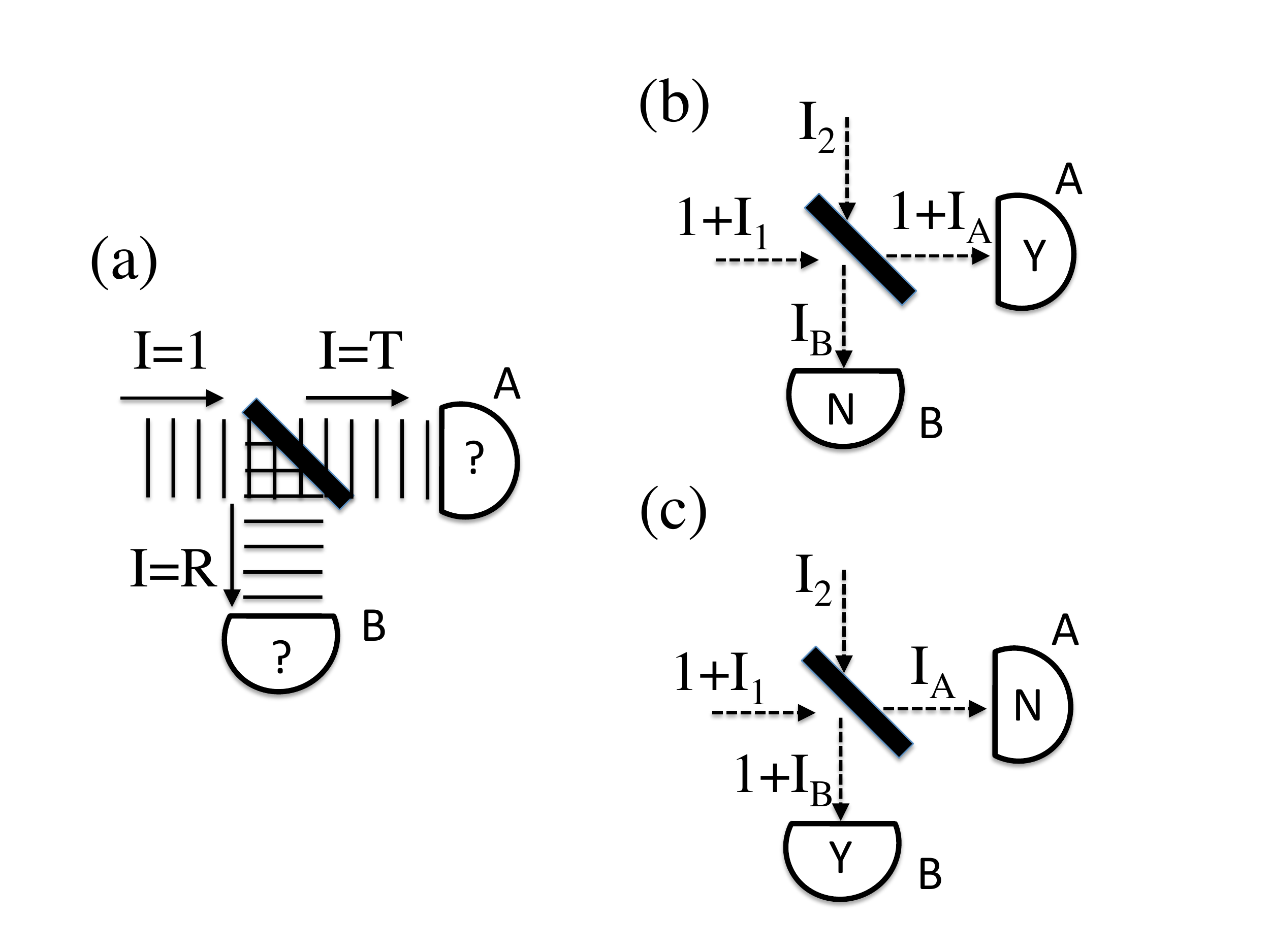}
\caption{(\textbf{a}) A classical photon analog encounters a beamsplitter, and is divided among two detectors, in contradiction with obervation. (\textbf{b}) A classical photon analog, boosted by some unknown peak intensity $I_1$, encounters the same beamsplitter.  Another beam with unknown peak intensity $I_2$ enters the dark port.  This is potentially consistent with a classical photon detection in only detector $A$ (``Y" for yes, ``N" for no), so long as the output intensities $I_A$ and $I_B$ remain unobserved. (The wavefronts have been replaced by dashed lines for clarity.)  (\textbf{c})  The same inputs as in Figure 1b, but with outputs consistent with classical photon detection in only detector $B$, where the output intensities $I_A$ and $I_B$ again remain unobserved.}
\end{figure} 

By adding these unknown input and output fields, Figure 1b demonstrates a classical beamsplitter scenario that is consistent with an observation of one CPA on detector $A$.  In this case two incoming beams, with peak intensities $1+I_1$ and $I_2$ interfere to produce two outgoing beams with peak intensities $1+I_A$ and $I_B$.  The four unknown intensities are related by energy conservation, $I_1+I_2=I_A+I_B$, where the exact relationship between these four parameters is determined by the unknown phase difference between the incoming beams.  Different intensities and phases could also result in the detection of exactly one CPA on detector $B$, as shown in Figure 1c.  These scenarios are allowed by classical EM and consistent with observation, subject to known uncertainties in measuring field values, pointing the way towards a classical account of ``single-photon'' experiments.  This is also distinct from prior field-based accounts of beamsplitter experiments \cite{kaloyerou2006}; here there is no need to non-locally transfer field energy from one path to another.

Some potential objections should be addressed.  One might claim that quantum theory \textit{does} allow certainty in the total energy of a photon, at the expense of timing and phase information.  But in quantum field theory one can only arrive at this conclusion after one has renormalized the zero-point values of the electromagnetic field -- the very motivation for $I_1$ and $I_2$ in the first place.\footnote{Furthermore, when hunting for some more-classical formulation of quantum theory, one should not assume that the original formulation is correct in every single detail.}  

Another objection would be to point out the sheer implausibility of any appropriate beam $I_2$.  Indeed, in order to interfere with the original CPA, it would have to come in with just the right frequency, spatial mode, pulse shape, and polarization.  But this concern makes the error of thinking of all past parameters as logical inputs.  In the Lagrangian Schema, the logical inputs are the known constraints at the beginning and end of the relevant system.  The unknown parameters are logical \textit{outputs} of this Schema, just as the initial angle of the light ray in Fermat's principle.  The below models will aim to \textit{generate} the parameters of the incoming beam in $I_2$, as constrained by the entire experiment.  In action principles, just because a parameter is coming into the system at the temporal beginning does not mean that it is a logical input.  In retrocausal models, these are the parameters that are the \textit{effects} of the constraints, not causes in their own right.\footnote{Such unknown background fields do not have external settings by which they can be independently controlled, even in principle, and therefore they are not causal interventions.}

Even if the classical field configurations depicted in Figures 1b and 1c are \textit{possible}, it still remains to explain why the observed transmission shown in Figure 1b occurs with a probability $T$, while the observed reflection shown in Figure 1c occurs with a probability $R$.  To extract probabilities from such a formulation, one obviously needs to assign probabilities to the unknown parameters, $P(I_1)$, $P(I_2)$, etc.  However, use of the Lagrangian Schema requires an important distinction, in that the probabilities an agent would assign to the unknown fields would depend on that agent's information about the experimental geometry.  In the absence of any information whatsoever, one would start with a ``\textit{a priori} probability distribution'' $P_0(I_2)$ -- effectively a Bayesian prior that would be (Bayesian) updated upon learning about any experimental constraints.  Any complete model would require both a probability distribution $P_0$ as well as rules for how the experimental geometry might further constrain the allowed field values.

Before giving an example model, one further problem should be noted.  Even if one were successful in postulating some prior distribution $P_0(I_1)$ and $P_0(I_2)$ that eventually recovered the correct probabilities, this might very well break an important time symmetry.  Specifically, the time-reverse of this situation would instead depend on $P_0(I_A)$ and $P_0(I_B)$.  For that matter, if both outgoing ports have a wave with a peak intensity of at least $I=1$, then the only parameters sensitive to which detector fires are the unobserved intensities $I_A$ and $I_B$.  Both of these arguments encourage us to include a consideration of the unknown outgoing intensities $I_A$ and $I_B$ in any model, not merely the unknown incoming fields.

\subsection{Simple Model Example}

The model considered in this section is meant to be an illustrative example of the class of retrocausal models described above, illustrating that it is possible to get particle-like phenomena from a field-based ontology, and also indicating a connection to some of the existing retrocausal accounts of entanglement.

One way to resolve the time-symmetry issues noted just above is to impose a model constraint whereby the two unobserved incoming intensities $I_1$ and $I_2$ are always exactly equal to the unobserved outgoing intensities $I_A$ and $I_B$.  (Either $I_1=I_A$ or $I_1=I_B$.)  If this constraint is enforced, then assigning a probability of $P_0(I_1)P_0(I_2)$ to each diagram does not break any time symmetry, as this quantity will always be equal to $P_0(I_A)P_0(I_B)$.  One simple rule that seems to work well in this case is the \textit{a priori} distribution
\begin{equation}
\label{eq:PI1}
P_0(I_Z)=Q\frac{1}{\sqrt{I_Z}} \;\; (where\,\, I_Z>\epsilon).
\end{equation}

Here $I_Z$ is any of the allowed unobserved background intensities, $Q$ is a normalization constant, and $\epsilon$ is some vanishingly small minimum intensity to avoid the pole at $I_Z=0$.  (While there may be a formal need to normalize this expression, there is never a practical need; these prior probabilities will be restricted by the experimental constraints before being utilized, and will have to be normalized again.)  The only additional rule to recover the appropriate probabilities is that $I_1>>\epsilon$.  (This might be motivated by the above analysis that laser photons would have to be triggered by background fields, so the known incoming CPA would have to be accompanied by a non-vanishing unobserved field.)

To see how these model assumptions lead to the appropriate probabilities, first consider that it is overwhelmingly probable that $I_2\approx\epsilon$.  So in this case we can ignore the input on the dark port of the beamsplitter.  But with only one non-vanishing input, there can be no interference, and both outputs must have non-vanishing intensities.  The only way it is possible for detector $A$ to fire, given the above constraints, is if $I_1=I_B=R/T$ in Figure 1b (such that $I_2=I_A=0$).  The only way it is possible for detector $B$ to fire, in Figure 1c, is if $I_1=I_A=T/R$.  

With this added information from the experimental geometry, one would update the prior distribution $P_0(I_1)$ by constraining the only allowed values of $I_1$ to be $R/T$ or $T/R$ (and then normalizing).  The relative probabilities of these two cases is therefore
\begin{equation}
\label{eq:calc1}
\frac{P(A)}{P(B)}=\frac{\frac{1}{\sqrt{R/T}}P_0(I_2)}{\frac{1}{\sqrt{T/R}}P_0(I_2)}=\frac{T}{R},
\end{equation}
yielding the appropriate ratio of possible outcomes.  

Taking stock of this result, here are the assumptions of this example model:  

\begin{itemize}[leftmargin=*,labelsep=5.8mm]
\item	The $\textit{a priori}$ probability distribution on each unknown field intensity is given by Eqn. (\ref{eq:PI1}) -- to be updated for any given experiment.
\item	The unknown field values are further constrained to be equal as pairs, $\{ I_1, I_2 \}=\{ I_A, I_B \}$.
\item $I_1$ is non-negligible because it accompanies a known ``photon''.
\item The probability of each diagram is given by $P_0(I_1)P_0(I_2)$, or equivalently, $P_0(I_A)P_0(I_B)$.
\end{itemize}  
 
Note that it does not seem reasonable to assign the prior probability to the total incoming field $(1+I_1)$, because (\ref{eq:PI1}) should refer to the probability given no further information, not even the knowledge that there is an incoming photon's worth of energy on that channel.  (The known incoming photon that defines this experiment is an addition to the \textit{a priori} intensity, not a part of it.)  Given these assumptions, one finds the appropriate probabilities for a detected transmission as compared to a detected reflection.

There are several other features of this example model.  Given Eqn. (\ref{eq:PI1}), it should be obvious that the total energy in most zero-point fields should be effectively zero, resolving the standard SED problem of infinite zero-point energy.  Also, this model would work for any device that splits a photon into two paths (such as a polarizing cube), because the only relevant parameters are the classical transmission and reflection, $T$ and $R$.

More importantly, this model allows one to recover the correct measurement probabilities for two maximally entangled photons in essentially the same way as several existing retrocausal models in the literature \cite{argaman2010,wharton2014,almada2015}.  Consider two  CPAs produced by parametric down-conversion in a nonlinear crystal, with identical but unknown polarizations (a standard technique for generating entangled photons).  The three-wave mixing that classically describes the down-conversion process can be strongly driven by the presence of background fields matching one of the two output modes, M1, even if there is no background field on the other output mode, M2.  (Given Eqn. (\ref{eq:PI1}), having essentially no background field on one of these modes is overwhelmingly probable.)  So in this case the polarization of M2 necessarily matches the polarization of the unknown background field on M1 (the field that strongly drives the down-conversion process).  

Now, assume both output photons are measured by polarizing cubes set at arbitrary polarization angles, followed by detectors.  With no extra background field on M2, the only way that M2 could satisfy the above constraints at measurement would be if its polarization was \textit{already} exactly aligned (modulo $\pi/2$) with the angle of the future polarizing cube.  (In that case, no background field would be needed on that path; the bare CPA would fully arrive at one detector or the other.)  But we have established that the polarization of M2 was selected by the background field on M1, so the background field on M1 is also forced to align with the measurement angle on M2 (modulo $\pi/2$).  In other words, solving the whole experiment ``all at once'', the polarization of both photons is effectively constrained to match one of the two future measurement angles.

This is essentially what happens in several previously-published retrocausal models of maximally entangled particles \cite{argaman2010, wharton2014, almada2015}.  In these models the properties of both particles (spin or polarization, depending on the context) are constrained to be aligned with one of the two future settings.  The resulting probabilities are then entirely determined by the mis-matched particle, the one doesn't match the future settings.  But this is just a single-particle problem, and in this case the corresponding classical probabilties (R and T, given by Malus's Law at the final polarizer) are enforced by the above rules, matching experimental results for maximally entangled particles.  The whole picture almost looks \textit{as if} the measurement on one photon has collapsed the other photon into that same polarization, but in these models it was clear that the CPAs had the correct polarization all along, due to future constraints on the appropriate hidden fields.

\subsection{Discussion}

The above model was presented as an illustrating example, demonstrating one way to resolve the most obvious problems with classical photon analogs and SED-style approaches.  Unfortunately, it does not seem to extend to more complicated situations.  For example, if one additional beamsplitter is added, as in Figure 2, no obvious time-symmetric extension of the assumptions in the previous section lead to the correct results.  In this case, one of the two dark ports would have to have non-negligible input fields.  Performing this analysis, it is very difficult to invent any analogous rules that lead to the correct distribution of probabilities on the three output detectors.

\begin{figure}[H]
\centering
\includegraphics[width=7cm]{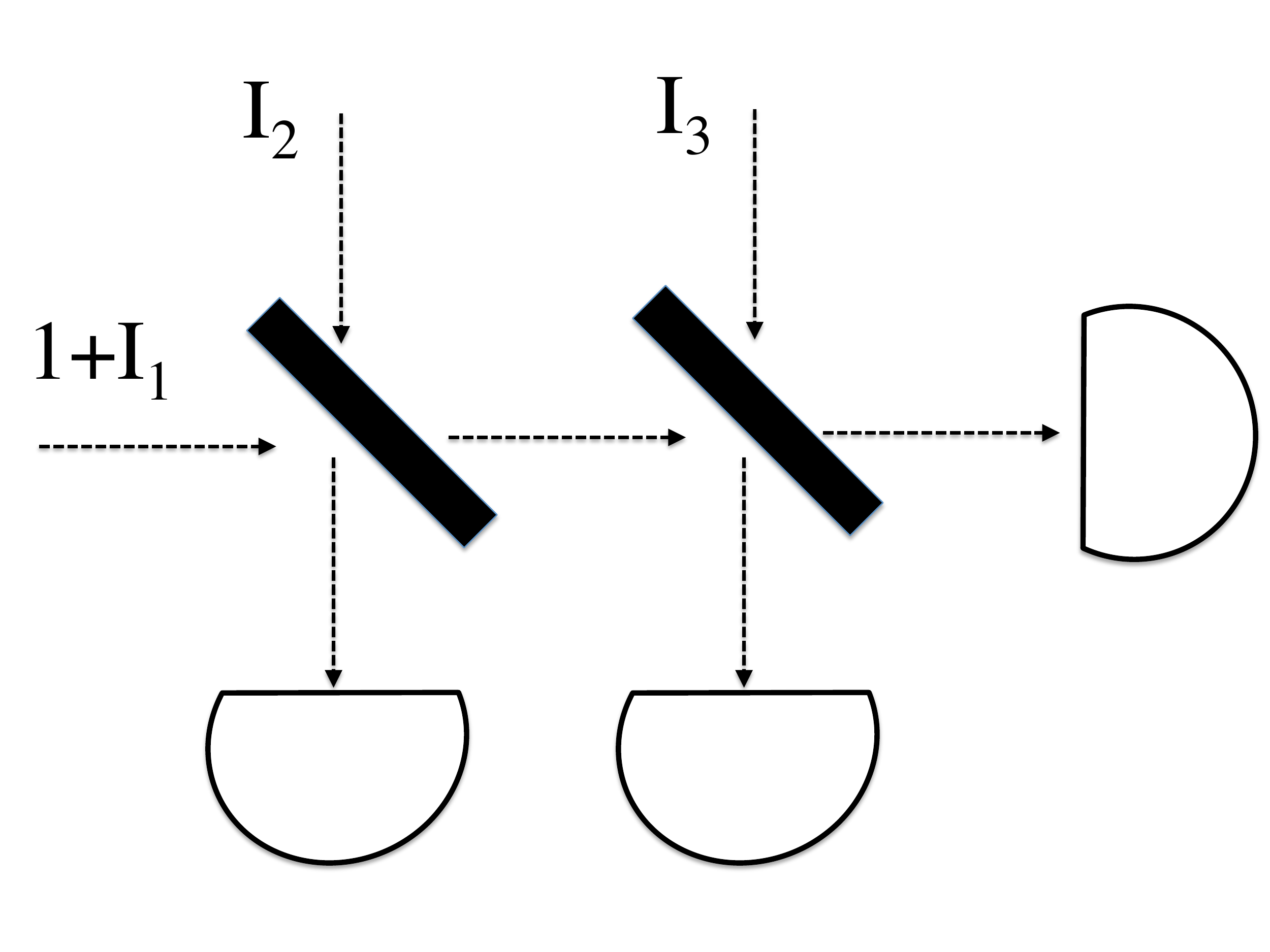}
\caption{(\textbf{a}) A classical photon analog encounters two beamsplitters, and is divided among three detectors.  The CPA is boosted by some unknown peak intensity $I_1$, and each beamsplitter's dark port has an additional incident field with unknown intensity.}
\end{figure} 

In Section 5 we will see that it is possible to resolve this problem, using different assumptions to arrive at another model which works fine for multiple beamsplitters.  But before proceeding, it is worth reviewing the most important accomplishment so far.  We have seen that it is possible to give a classical field account of an apparent single photon passing through a beamsplitter, matching known observations.  Such models are generally thought to be impossible (setting aside nonlocal options \cite{kaloyerou2006}). Given that they \textit{are} possible -- if using the Lagrangian Schema -- the next-level concern could be that such models are simply \textit{implausible}.  For phenomena that look so much like particle behavior, such classical-field-based models might seem to be essentially unmotivated. 

The next section will try to address this concern in two different ways.  First, the experiments considered in Section 4 will be expanded to include clear wave-like behavior, by combining two beamsplitters into an interferometer.  Again, the input and output will look like single particles, but now some essential wave interference is clearly occurring in the middle.  Second, the averaged and post-selected results of these models can be compared with ``weak values'' that can be measured in actual experiments.\cite{AAV,dressel2014a}  Notably, the results will demonstrate a new connection between the average intermediate classical fields and experimental weak values.  This correspondence is known in the high-field case \cite{dressel2014b,dressel2015,CWV1,CWV2,CWV3}, but here they are shown to apply even in the single-photon regime.  Such a result will boost the general plausibility of this classical-field-based approach, and will also motivate an improved model for Section 5. 

\section{Averaged Fields and Weak Values}

Even without a particular retrocausal model, it is still possible to draw conclusions as to the long-term averages predicted over many runs of the same experiment.  The only assumption made here will be that every relevant unknown field component for a given experiment (both inputs and outputs) is treated the same as every other.  In Figures 1a and 1b this would imply an equality between the averaged values $<\!I_1\!>=<\!I_2\!>=<\!I_A\!>=<\!I_B\!>$, each defined to be the quantity $I_Z$. 

Not every model will lead to this assumption; indeed, the example model above does not, because the the CPA-accompaning field $I_1$ was treated differently from the dark port field $I_2$.  But for models which do not treat these fields differently, the averages converge onto parameters that can actually be measured in the laboratory: weak values \cite{AAV,dressel2014a}.  This intriguing correspondence is arguably an independent motivation to pursue this style of retrocausal models.

\subsection{Beamsplitter Analysis}

Applying this average condition on the simple beamsplitter example of Figure 1b and 1c yields a phase relationship between the incoming beams, in order to retain the proper outputs.  If $\theta$ is the phase difference between $I_1$ and $I_2$ before the beamsplitter, then taking into account the relative $\pi/2$ phase shift caused by the beamsplitter itself, a simple calculation for Figure 1b reveals that
\begin{eqnarray}
\label{eq:IA1}
\left< 1+I_A\right>=&I_Z+T-\left<2\sqrt{RT (1+I_1)(I_2)}\sin\theta\right> \\
\label{eq:IB1}
\left< I_B\right>=&I_Z+R+\left<2\sqrt{RT (1+I_1)(I_2)}\sin\theta\right>.
\end{eqnarray}

Given the above restrictions on the average values, this is only possible if there exists a non-zero average correlation
\begin{equation}
\label{eq:C}
C\equiv\left<\sqrt{(1+I_1)(I_2)}\sin\theta\right>
\end{equation}
between the inputs, such that $C=-\sqrt{R/4T}$.  The same analysis applied to Figure 1c reveals that in this case $C=\sqrt{T/4R}$.  (This implies some inherent probability distribution $P(I_1,I_2,\theta)\propto 1/|C|$ to yield the correct distribution of outcomes, which will inform some of the model-building in the next section.)  In this case, there are no intermediate fields to analyze, as every mode is either an input or an output.  To discuss intermediate fields, we must go to a more complicated scenario.

\subsection{Interferometer Analysis}

Consider the simple interferometer shown in Figure 3.  For these purposes, we will assume it is aligned such that the path length on the two arms is exactly equal.  For further simplicity, the final beamsplitter is assumed to be 50/50.  Again, the global constraints imply that either Figure 3a or Figure 3b actually happens.  A calculation of the average intermediate value of $I_x$ yields the same result as Eqn. (\ref{eq:IA1}), while the average value of $I_y$ is the same as Eqn. (\ref{eq:IB1}).  For Figure 3a, further interference at the final beamsplitter then yields, after some simplifying algebra,
\begin{eqnarray}
\label{eq:IA2}
\left< 1+I_A\right>=&(0.5+\sqrt{RT})+I_Z+(T-R)\left<\sqrt{(1+I_1)(I_2)}\sin\theta\right> \\
\label{eq:IB2}
\left< I_B\right>=&(0.5-\sqrt{RT})+I_Z-(T-R)\left<\sqrt{(1+I_1)(I_2)}\sin\theta\right>.
\end{eqnarray}

The first term on the right of these expressions is the outgoing classical field intensity one would expect for a single CPA input, with no unknown fields.  Because of our normalization, it is also the expected probability of a single-photon detection on that arm.  The second term is just the average unknown field $I_Z$, and the final term is a correction to this average that is non-zero if the incoming unknown fields are correlated.  Note that the quantity $C$ defined in Eqn. (\ref{eq:C}) again appears in this final term.  

\begin{figure}[H]
\centering
\includegraphics[width=10 cm]{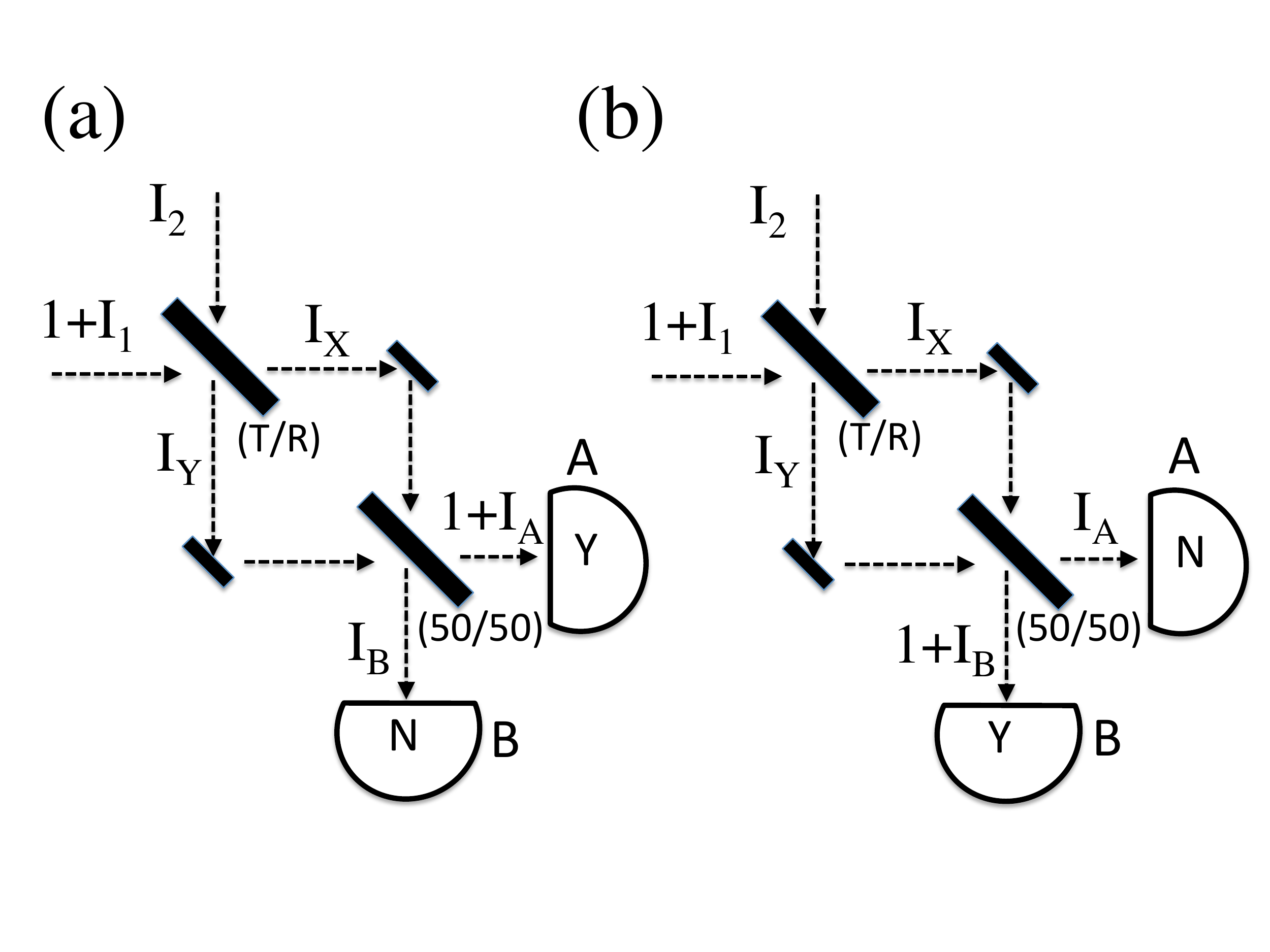}
\caption{(\textbf{a}) A classical photon analog, boosted by some unknown peak intensity $I_1$, enters an interferometer through a beamsplitter with transmission fraction $T$.  An unknown field also enters from the dark port.  Both paths to the final 50/50 beamsplitter are the same length; the intermediate field intensities on these paths are $I_X$ and $I_Y$.  Here detector $A$ fires, leaving unmeasured output fields $I_A$ and $I_B$.   (\textbf{b}) The same situation as Figure 3a, except here detector $B$ fires.  }
\end{figure} 

In order to make this end result compatible with the condition that $\left< 1+I_A\right>=1+ I_Z$, the correlation term $C$ must be constrained to be $C\!=\!(0.5-\sqrt{RT})/(T-R)$.  For Figure 3b, with detector $B$ firing, this term must be $C\!=\!-(0.5+\sqrt{RT})/(T-R)$.  (As in the beamsplitter case, the quantity $1/|C|$ happens to be proportional to the probability of the corresponding outcome, for allowed values of $C$.)  Notice that as the original beamsplitter approaches 50/50, the required value of $C$ diverges for Figure 3b, but not for Figure 3a.  That is because this case corresponds to a perfectly tuned interferometer, where detector $A$ is certain to fire, but never $B$.\footnote{This analysis also goes through for an interferometer with an arbitrary phase shift, and arbitrary final beamsplitter ratio; these results will be detailed in a future publication.}  

In this interferometer, once the outcome is known, it is possible to use $C$ to calculate the average intensities $<\!I_X\!>$ and $<\!I_Y\!>$ on the intermediate paths.  For Figure 3a, some algebra yields:
\begin{eqnarray}
\label{eq:Ix1}
\left< I_X\right>=&I_Z+\frac{\sqrt{T}}{\sqrt{T}+\sqrt{R}} \\
\label{eq:Iy2}
\left< I_Y\right>=&I_Z+\frac{\sqrt{R}}{\sqrt{T}+\sqrt{R}}.
\end{eqnarray}

For Figure 3b, the corresponding average intermediate intensities are
\begin{eqnarray}
\label{eq:Ix2}
\left< I_X\right>=&I_Z+\frac{\sqrt{T}}{\sqrt{T}-\sqrt{R}} \\
\label{eq:Iy2}
\left< I_Y\right>=&I_Z-\frac{\sqrt{R}}{\sqrt{T}-\sqrt{R}}.
\end{eqnarray}
Remarkably, as we are about to see, the non-$I_Z$ portion of these calculated average intensities can actually be measured in the laboratory.

\subsection{Weak Values}

When the final outcome of a quantum experiment is known, it is possible to elegantly calculate the (averaged) result of a weak intermediate measurement via the real part of the ``Weak Value'' equation \cite{AAV}:
\begin{equation}
\label{eq:WV}
\left< \bf{Q} \right>_{weak} = Re \left( \frac {\bra{\Phi} {\bf Q} \ket{\Psi}}{\ip{\Phi}{\Psi}} \right).
\end{equation}
Here $\ket{\Psi}$ is the initial wavefunction evolved forward to the intermediate time of interest, $\ket{\Phi}$ is the final (measured) wavefunction evolved backward to the same time, and $\bf{Q}$ is the operator for which one would like to calculate the expected weak value.\footnote{Note that weak values by themselves are not retrocausal.  Post-selecting an outcome is not a causal intervention.  (One cannot ``cause'' a trip to be faster by post-selecting on green traffic lights.)  And without post-selection, the averaged weak values show no correlation with the future setting.  (However, if one takes the backward-evolved wavefunction $\ket{\Phi}$ to be an element of reality, as done by one of the authors here \cite{aharonov2008}, then one does have a retrocausal model -- albeit in configuration space rather than spacetime.)}  This equation yields the correct answer in the limit that the measurement $\bf{Q}$ is sufficiently weak, so that it does not appreciably affect the intermediate dynamics.  The success of this equation has been verified in the laboratory \cite{CWV1}, but is subject to a variety of interpretations.  For example, $\left< \bf{Q} \right>_{weak}$ can be negative, seemingly making a classical interpretation impossible.

In the case of the interferometer, the intermediate weak values can be calculated by recalling that it is the square root of the normalized intensity that maps to the wavefunction.  (And of course, the standard wavefunction knows nothing about $I_Z$; only the prepared and detected photon are relevant in a quantum context.)  Taking into account the phase shift due to a reflection, the wavefunction between the two beamsplitters is simply $\ket{\Psi}=\sqrt{T}\ket{X}+i\sqrt{R}\ket{Y}$, where $\ket{X} (\ket{Y})$ is the state of the photon on the upper (lower) arm of the interferometer.  

The intermediate value of $\ket{\Phi}$ depends on whether the photon is measured by detector $A$ or $B$. The two possibilities are:
\begin{eqnarray}
\label{eq:phia}
\ket{\Phi_A}&=&\frac{1}{\sqrt{2}} \left( -i\ket{X}+\ket{Y} \right),\\
\label{eq:phib}
\ket{\Phi_B}&=&\frac{1}{\sqrt{2}} \left( \ket{X}-i\ket{Y} \right) .
\end{eqnarray}
Notice that in this case the reflection off the beamsplitter is associated with a negative $\pi/2$ phase shift, because we are evolving the final state in the opposite time direction.

These are easily inserted into Eqn. (\ref{eq:WV}), where ${\bf Q}=\ket{X}\bra{X}$ for a weak measurement of $I_X$, and ${\bf Q}=\ket{Y}\bra{Y}$ for a weak measurement of $I_Y$ . (Given our normalization, probability maps to peak intensity.)  If the outcome is a detection on $A$, this yields
\begin{eqnarray}
\label{eq:Ix3}
\left< I_X\right>_{weak} =&\frac{\sqrt{T}}{\sqrt{T}+\sqrt{R}}, \\
\label{eq:Iy3}
\left< I_Y\right>_{weak} =&\frac{\sqrt{R}}{\sqrt{T}+\sqrt{R}}.
\end{eqnarray}
If instead the outcome is a detection on $B$, one finds
\begin{eqnarray}
\label{eq:Ix4}
\left< I_X\right>_{weak} =&\frac{\sqrt{T}}{\sqrt{T}-\sqrt{R}}, \\
\label{eq:Iy4}
\left< I_Y\right>_{weak} =&\frac{-\sqrt{R}}{\sqrt{T}-\sqrt{R}}.
\end{eqnarray}
Except for the background average intensity $I_Z$, these quantum weak values are precisely the same intermediate intensities computed in the previous section.  

The earlier results were framed in an essentially classical context, but these weak values come from an inherently quantum calculation, with no clear interpretation.  Some of the strangest features of weak values are when one gets a negative probability/intensity, which seem to have no classical analog whatsoever.  For example, whenever detector $B$ fires, either (\ref{eq:Ix4}) or (\ref{eq:Iy4}) will be negative.  (Recall that if $T=R$, then $B$ never fires.)    Nevertheless, a classical interpretation of this negative weak value is still consistent with the earlier results of (\ref{eq:Ix2}) and (\ref{eq:Iy2}), because those cases also include an additional unknown intensity $I_Z$.  It is perfectly reasonable to have classical destructive interference that would decrease the average value of $I_Y$ to below that of $I_Z$; after all, the latter is just an unknown classical field.  

One objection here might be that for values of $T\approx R$, the weak values (\ref{eq:Ix4}) and (\ref{eq:Iy4}) could get arbitrarily large, such that $I_Z$ would have to be very large as well to maintain a positive intensity for both (\ref{eq:Ix2}) and (\ref{eq:Iy2}).  But consider that if $I_Z$ were \textit{not} large enough, then there would be no classical solution at all, in contradiction to the Lagrangian Schema assumptions considered above (requiring a global solution to the entire problem).  Furthermore, if the weak values get very large, that is only because the outcome at $B$ becomes very improbable, meaning that $I_Z$ would rarely have to take a large value.  As we shall see in the next section, there are reasonable \textit{a priori} distributions of $I_Z$ which would be consistent with this occasional restriction.

Such connections between uncertain classical fields and quantum weak values are certainly intriguing, and also under current investigation by at least one other group \cite{steinberg2018}.  But while it may be that the unknown-classical-field framework might help make some conceptual sense of quantum weak values, the main point here is simply that these two perspectives are mutually \textit{consistent}.  Specifically, the known experimental success of weak value predictions seems to equally support the unknown-field formalism presented above.  It remains to be seen whether (and why) these two formalisms always seem to give compatible answers in every case, but this paper will set that question aside for future research.

For the purposes of this introductory paper, the final task will be to consider whether the above results indicate a more promising model of these experiments.

\section{An Improved Model}

Given the intriguing connection to weak values demonstrated in the previous section, it seems worth trying to revise the example model from Section 3.  In Section 4, the new assumption which led to the successful result was that every unknown field component $(I_1,I_2,I_A,I_B)$, should be treated on an equal footing, not singling out $I_1$ for accompanying a known photon.  (Recall the average value of each of these was assumed to be some identical parameter $I_Z$.)  Meanwhile, the central idea from the model in Section 3 was that time-symmetry could be enforced by demanding an exact equivalence between the two input fields ($I_1,I_2$) and the two output fields ($I_A,I_B$).

One obvious way to combine all these ideas is to instead demand an equivalence between all \textit{four} of these intensities -- not on average, but on every run of the experiment.  This might seem to be in conflict with the weak value measurements, which are not the same on every run, but only converge to the weak values after an experimental averaging.  However, these measurements are necessarily weak/noisy, so these results are inconclusive as to whether the underlying signal is constant or varying. (Alternatively, one could consider a class of models that on average converge to the below model, but this option will also be set aside for the purposes of this paper.)

With the very strict constraint that each of  ($I_1,I_2,I_A,I_B$) are always equal to the same intensity $I_Z$, the only two free parameters are $I_Z$ and the relative initial phase $\theta$ (between the two incoming modes $1+I_1$ and $I_2$).  Also, $\theta$ and $I_Z$ must be correlated, depending on the experimental parameters, in order to fulfill these constraints.  For the case of the beamsplitter (Figures 1b and 1c), this amounts to removing all the time-averages from the analysis of Section 4.1.  This leads to the conditions
\begin{eqnarray}
\label{eq:thetaa}
\frac{1}{\sqrt{I_{ZA}^2 + I_{ZA}}}&=&-\sin{\theta} \sqrt{\frac{4T}{R}},\\
\label{eq:thetab}
\frac{1}{\sqrt{I_{ZB}^2+I_{ZB}}}&=&\sin{\theta} \sqrt{\frac{4R}{T}}.
\end{eqnarray}

Here $I_{ZA}$ is the value of $I_Z$ needed for an outcome on detector $A$ (as in Figure 1b), and $I_{ZB}$ is the value of $I_Z$ needed for an outcome on detector $B$ (as in Figure 1c).  Both of these are functions of $\theta$.

This model requires \textit{a priori} probability distributions $P_0(I_Z)$ and $P'_0(\theta)$ (the prime is to distinguish these two functions).  The hope is that these distributions can then be restricted by the global constraints such that the correct outcome probabilities are recovered.  To implement the above constraints, instead of integrating over the two-dimensional space [$I_Z,\theta$], the correlations between $I_Z$ and $\theta$ essentially make this a one-dimensional space, which can be calculated with a delta function:
\begin{equation}
\label{eq:cond1}
\frac{\int P_0(I_Z) P'_0(\theta) \delta(I_Z-I_{ZA}) dI_Z d\theta}{\int P_0(I_Z) P'_0(\theta) \delta(I_Z-I_{ZB}) dI_Z d\theta} = \frac {P(outcome\,A)}{P(outcome\,B)}.
\end{equation}

It is very hard to imagine any rule whereby $P'_0(\theta)$ would not start out as a flat distribution -- all relative phases should be equally \textit{a priori} likely.  The earlier observation that the appropriate probability was always proportional to $1/|C|$ (in both the beamsplitter and the interferometer geometries) motivates the following guess for an \textit{a priori} probability distribution for background fields:
\begin{equation}
\label{eq:money}
P_0(I_Z) \propto \frac{1}{\sqrt{I_Z^2 + I_Z}},
\end{equation}
assuming the normalization where $I=1$ corresponds to a single classical photon.  This expression diverges as $I_Z\to 0 $, which is appropriate for avoiding the infinities of SED, although some cutoff would be required to form a normalized distribution.\footnote{Again, it is unclear whether an \textit{a priori} assessment of relative likelihood would actually have to be normalized, given that in any experimental instance there would only be some values of $I_Z$ which were possible, and only these probabilities would have to be normalized.}

Inserting (\ref{eq:money}) into (\ref{eq:cond1}), along with a flat distribution for $P'_0(\theta)$, the beamsplitter conditions from (\ref{eq:thetaa}) and (\ref{eq:thetab}) yield
\begin{equation}
\label{eq:cond2}
\frac{\int_\pi^{2\pi} -sin\theta \sqrt{4T/R} d\theta}{\int_0^\pi  sin\theta \sqrt{4R/T} d\theta} = \frac {T}{R},
\end{equation}
as desired.  Here the limits on $\theta$ come from the range of possible solutions to Eqns. (\ref{eq:thetaa}) and (\ref{eq:thetab}).  A similar successful result is found in the above case of the interferometer, because $1/|C|$ is again proportional to the outcome probability.  This model also works well for the previously-problematic case of multiple beamsplitters shown in Figure 2.  Now, because the incoming fields $(I_1,I_2,I_3)$ are all equal, this essentially splits into two consecutive beamsplitter problems, and the probabilities of these two beamsplitters combine in an ordinary manner. 

Summarizing the assumptions behind this improved model:

\begin{itemize}[leftmargin=*,labelsep=5.8mm]
\item	The unknown field values are constrained to all be equal; $I_1=I_2 =I_A=I_B$.
\item	The $\textit{a priori}$ probability distribution on each unknown field intensity is given by Eqn. (\ref{eq:money}) -- but must be updated for any given experiment.
\item The relative phase between the incoming fields is \textit{a priori} completely unknown -- but must be updated for any given experiment.
\end{itemize} 

However, there is still a conceptual difficulty in this new model, in that all \textit{considered} incoming field modes are constrained to be equal intensities, but we have left the \textit{unconsidered} modes equal to zero.  (Meaning, the modes with the wrong frequencies, or coming in the wrong direction, etc.)  If literally \textit{all} zero-point modes were non-zero, it would not only change the above calculations, but it would run directly into the usual infinities of SED.  So if this improved model were to be further developed, there would have to be some way to determine certain groups of background modes that were linked together through the model assumptions, while other background modes could be neglected.  

This point is also essential if such a revised model is to apply to entangled particles.  For two down-converted photons with identical polarizations, each measured by a separate beamsplitter, there are actually 4 relevant incoming field modes: the unknown intensity accompanying each photon, as well as the unknown intensity incident upon the dark port of each beamsplitter.  If one sets all 4 of these peak intensities to the same $I_Z$, one does not recover the correct joint probabilities of the two measurements.  But if two of these fields are (nearly) zero, as described in Section 3.2, then the correct probabilities are recovered in the usual retrocausal manner (see Section 3.2 or \cite{argaman2010,wharton2014,almada2015}).  Again, it seems that there must be some way to parse the background modes into special groups. 

The model in this section is meant to be an example starting point, not some final product.  Additional features and ideas that might prove useful for future model development will now be addressed in the final section.

\section{Summary and Future Directions}

Retrocausal accounts of quantum phenomena have come a long way since the initial proposal by Costa de Beauregard \cite{costa1953}.  Notably, the number of retrocausal models in the literature has expanded significantly in the past decade alone \cite{argaman2010,wharton2010b,wharton2011,evans2012,harrison2012,schulman2012,heaney2013,corry2015,lazarovici2015,sutherland2017,price2008,wharton2014,almada2015,weinstein2017,silberstein2018,sen2018}, but more ideas are clearly needed.  The central novelties in the class of models discussed here are (1) using fields (exclusively) rather than particles, and (2) introducing uncertainty to even the initial and final boundary constraints.  Any retrocausal model must have hidden variables (or else there is nothing for the future measurement choices to constrain), but it has always proved convenient to segregate the known parameters from the unknown parameters in a clear manner.  Nature, however, may not respect such a convenience.  In the case of realistic measurements on fields, there is every reason to think that our best knowledge of the field strength may not correspond to the actual value.

Although the models considered here obey classical field equations (in this case, classical electromagnetism), they only make sense in terms of the Lagrangian Schema, where the entire experiment is solved ``all-at-once''.  Only then does it make sense to consider incoming dark-port fields (such as $I_2$), because the global solution may require these incoming modes in order have a solution.  But despite the presence of such fields at the beginning of the experiment (and, presumably, before it even begins), they are not ``inputs'' in the conventional sense; they are literally outputs of the retrocausal model.

The above models have demonstrated a number of features and consequences, most notably:
\begin{itemize}[leftmargin=*,labelsep=5.8mm]
\item	Distributed classical fields can be consistent with particle-like detection events.
\item There exist simple constraints and \textit{a priori} field intensity distributions that yield the correct probabilities for basic experimental geometries.
\item	Most unobserved field modes are expected to have zero intensity (unlike in SED).
\item	The usual retrocausal account for maximally entangled photons still seems to be available.
\item The average intermediate field values, minus the unobserved background, is precisely equal to the ``weak value'' predicted by quantum theory (in the cases considered so far).
\item Negative weak values can have a classical interpretation, provided the unobserved background is sufficiently large.
\end{itemize}

This seems to be a promising start, but there are many other research directions that might be inspired by these models.  For example, consider the motivation of action constraints, raised in Section 2.  If the total action is ultimately important, then any constraint or probability rule would have to consider the contribution to the action of the microscopic intermediate fields.  Even the simple case of a CPA passing through a finite-thickness beamsplitter has a non-trivial action. (A single free-field EM wave has a vanishing Lagrangian density at every point, but two crossing or interfering waves generally do not).  It certainly seems worth developing models that constrain not only the inputs and outputs, but also these intermediate quantities (which would have the effect of further constraining the inputs and outputs).  

Another possibility is to make the incoming beams more realistic, introducing spatially-varying noise, not just a single unknown parameter per beam.  It is well-known that such spatial noise introduces bright speckles into laser profiles, and in some ways these speckles are analogous to detected photons -- both in terms of probability distributions as well as their small spatial extent (compared to the full laser profile).  A related point would be to introduce unknown \textit{matter} fields, say some zero-point equivalent of the classical Dirac field, which would introduce further uncertainty and effective noise sources into the electromagnetic field.  These research ideas, and other related approaches, are wide open for exploration.

Certainly, there are also conceptual and technical problems that need to be addressed, if such models are to be further developed.  The largest unaddressed issue is how a global action constraint applied to macroscopic measurement devices might lead to specific rules that constrain the microscopic fields in a manner consistent with observation.  (In general, two-time boundary constraints can be shown to lead to intermediate particle-like behavior \cite{wharton2010a}, but different global rules will lead to different intermediate consequences.)  The tension between a covariant action and the special frame of the measurement devices also needs to be treated consistently.  Another topic that is in particular need of progress is an extension of retrocausal entanglement models to handle partially-entanged states, and not merely the maximally-entangled Bell states.  

Although the challenges remain significant, the above list of accomplishments arising from this new class of models should give some hope that further accomplishments are possible.  By branching out from particle-based models to field-based models, novel research directions are clearly motivated.  The promise of such research, if successful, would be to supply a nearly-classical explanation for all quantum phenomena: realistic fields as the solution to a global constraint problem in spacetime.

\acknowledgments{The author would like to thank J. Dressel for very helpful advice, A. Steinberg for unintentional inspiration, J. Walleczek for crucial support and encouragement, R. Bahuguna for insights concerning laser speckles, and M. Leifer for hosting a productive visit to Chapman University.  This work is supported in part by the Fetzer Franklin Fund of the John E. Fetzer Memorial Trust.}


\externalbibliography{yes}
\bibliography{EMQMbib}

\end{document}